# An Empirical Study on Decision-Making Aspects in Responsible Software Engineering for AI

Lekshmi Murali Rani, Faezeh Mohammadi, Robert Feldt, Richard Berntsson Svensson
*Department of Computer Science and Engineering*
*Chalmers University of Technology and University of Gothenburg*
SE - 41296, Gothenburg, Sweden
Emails: lekshmi@chalmers.se, faezehm@chalmers.se, robert.feldt@chalmers.se, richard@cse.gu.se

*Abstract*—Incorporating responsible practices into software engineering for AI is essential to ensure ethical principles, societal impact, and accountability remain at the forefront of AI system design and deployment. This study investigates the ethical challenges and complexities inherent in responsible software engineering for AI, underscoring the need for practical, scenario-driven operational guidelines. Given the complexity of AI and the relative inexperience of professionals in this rapidly evolving field, continuous learning and market adaptation are crucial. Through qualitative interviews with seven practitioners (conducted until saturation), quantitative surveys of 51 practitioners and static validation of results with four industry experts in AI, this study explores how personal values, emerging roles, and awareness of AI's societal impact influence responsible decision-making in responsible software engineering for AI. A key finding is the gap between the current state of the art and actual practice in responsible software engineering for AI, particularly in the failure to operationalize ethical and responsible decision-making within the software engineering life cycle for AI. While ethical issues in responsible software engineering for AI largely mirror those found in broader software engineering process, the study highlights a distinct lack of operational frameworks and resources to guide responsible software engineering practices for AI effectively. The results reveal that current ethical guidelines are insufficiently implemented at the operational level, reinforcing the complexity of embedding ethics throughout the software engineering life cycle. The study concludes that interdisciplinary collaboration, H-shaped competencies (Ethical-Technical dual competence), and a strong organizational culture of ethics are critical for fostering responsible software engineering practices for AI, with a particular focus on transparency and accountability.

## I. INTRODUCTION

The extensive development of AI software exposes organizations and society to new challenges and risks, including data privacy breaches, algorithmic bias, accountability gaps, and lack of transparency [1]. Managing AI-related risks requires systematically identifying, assessing, and addressing their root causes that usually emerge at different stages of the AI system's lifecycle [2]. While studies emphasize the need to make informed, smart, value-based decisions to ensure critical success in software engineering (software engineering) practices [3], they do not provide any method to understand the influence of responsible decision-making in building responsible software. This highlights the need to study the decision-making process involved in developing these systems to understand the complexities of their impact on individuals and society. Responsible software engineering (RSE) for AI (also known as responsible AI engineering) aims to identify the challenges that practitioners face in developing AI systems responsibly and tries to mitigate these risks by ensuring adherence to ethical guidelines and software engineering practices for RAI. It incorporates ethical considerations, accountability measures, transparency, and fairness in decision-making processes, thereby reducing potential adverse outcomes. Software engineering serves as the foundation for implementing RAI by providing guidelines and tools for various stakeholders [4] and addressing risks that arise from AI's complexity and unpredictability [5]. The incorporation of responsibility and ethics in AI adds potential issues as software engineers prioritize quality and safety, while business managers prioritize revenue generation [6]. These ongoing issues are amplified by the popularity of AI, which may have long-term consequences and influence human behavior.

Responsible AI principles cover privacy preservation, accountability, safety, security, transparency, fairness, explainability, human control of technology, and promotion of human values [4]. The objective of these principles is to ensure that AI systems behave and make decisions responsibly, taking into account both legal and ethical requirements. Decision-making in responsible AI engineering involves balancing trade-offs and aligning choices with ethical considerations and societal impact [7]. The successful and responsible implementation of any software can be traced back to the decisions that were made at an earlier point of time in the software engineering life cycle [8]. It is therefore essential to examine the decision-making process in responsible AI engineering teams, focusing on the unique factors in RAI practices. While software engineers are crucial in any software development scenario, the new emerging roles (data engineers, AI ethicists, data scientists, AI engineers) introduce new and diverse sets of values and perspectives that are critical in operationalizing ethical principles in an AI system development life cycle. Understanding the cognitive, behavioral, and social aspects of these roles is vital in software engineering scenarios. This holistic view, known as Behavioral Software Engineering [9], is important for exploring human aspects in responsible AI engineering. The influence of emerging roles in software teams and their broader perspectives on decision-making should also be studied by further understanding the critical aspects of decision-making in responsible AI engineering and how it affects software development teams and processes.

To understand the decision-making process in the responsible AI engineering and the relevance of emerging roles and human factors in these decision-making process, an exploratory knowledge-seeking research was conducted using mixed method approach. The study involved 1 pilot interview followed by data collected through in-depth semi-structured interviews with seven industry practitioners, a questionnaire with 51 respondents, and a static validation of the results using semi-structured interviews with four industry practitioners. The obtained results underwent thematic analysis and narrative analysis, followed by a static validation of findings with the help of industry experts. The study revealed the relevance of interdisciplinary collaboration, dual-competence in ethical and technical aspects, the influence of individual and organizational values on decision-making, and the need for structured ethical guidelines and frameworks in RAI development. These findings and their implications are discussed in this paper.

The remainder of this paper is organized as follows. Section 2 presents related work. Section 3 describes the design of the mixed-method study. Section 4 presents the results and analysis, while the results are discussed in Section 5. Section 6 discloses the threats to the validity of the study. Finally, Section 7 gives a summary of the main conclusions.

## II. RELATED WORK

The advancement and evolution in AI calls for RAI implementation, which prioritizes transparency, accountability, inclusivity, fairness, ethics, managing data for privacy and security [6], [10]–[12]. The current landscape of RAI involves addressing the social and ethical risks associated with AI technologies but responsible AI engineering faces technical, legal and social challenges such as safety, privacy, biases, and societal impacts [13] that require a multifaceted approach, involving technical advancements, regulatory measures, and international cooperation to ensure AI contributes positively to society while minimizing potential harms. A roadmap for software engineering for RAI is essential to address challenges faced by stakeholders in meeting RAI standards [4]. Recommendations include governance structures, diverse participation, and industry-supported research centers [14]. Researchers highlight the difficulties in operationalizing ethical AI principles and advocate for responsible design, development, and validation [15]. Jobin et al. [16] advocate for the global convergence of ethical AI principles in AI guideline development. The black box problem in AI systems complicates transparency, making explainable AI frameworks essential for clarity in decision-making [17]. As technology companies are investing in RAI to increase algorithmic accountability, they face challenges in balancing innovation with ethical considerations [6], [10], [18]. Researchers emphasize the importance of RSE, highlighting the critical need for trustworthiness, quality, and ethical usage in guiding the development of socially responsible software [19].

Decision-making is a critical component in responsible AI development, which involves balancing the trade-offs and making choices that align with ethical considerations, the impact on society, and technical feasibility [7]. Decision-making in software engineering has evolved from short-term considerations to value-based approaches, focusing on long-term value creation [3]. Gogoll et al. [20] emphasize the importance of ethical decision-making at strategic, tactical, and operational levels, highlighting the limitations of traditional Codes of Conduct in guiding complex ethical situations. Effective decision-making requires a systematic approach considering technical feasibility, cost, schedule, risk, and ethical implications [20]. MacNamara et al. [21] conducted an empirical study on the influence of the ACM Code of Ethics on software engineers' decision-making, revealing challenges in applying abstract principles to real-world situations and the interplay between personal values and professional ethics.

As we go deeper into the landscape of RAI, it becomes apparent that the field extends beyond mere technical rectifications to include profound ethical and behavioral considerations. This realization brings us to the significance of Behavioral Software Engineering and its role in understanding the human factors that influence responsible AI development. Behavioral Software Engineering investigates cognitive, behavioral, and social aspects of software engineering, crucial for understanding human factors in ethical AI development [22]. Behavioral Software Engineering principles contribute to RAI by integrating human factors like motivations and values of individuals [23], [24], aiding in understanding and addressing ethical complexities in AI development. This underscores the importance of aligning AI systems with human values [25]. Overall, understanding motivation, values, and ethics in software engineering is essential for fostering a productive and ethically responsible development environment [3], [26].

The existing literature emphasizes the relevance of ethical considerations but overlooks how these principles can be operationalised within the decision-making process. This study aims to bridge the gap in existing research by offering a comprehensive analysis of the ethical considerations and decision-making processes in RAI development when compared to conventional software development (CSD). It enriches the field by highlighting the influence of human and social dimensions of responsible AI engineering and advocating for a nuanced understanding of the philosophical foundations and varied research paradigms that shape AI development practices.

## III. RESEARCH METHODOLOGY

This section discusses the research methodology used in this study, including the research design, research questions, data collection, and analysis process.

### A. Research Design

The research design for this study is based on a decision-making structure (Figure 1) to ensure that well-founded and informed choices are made throughout the research process [27].

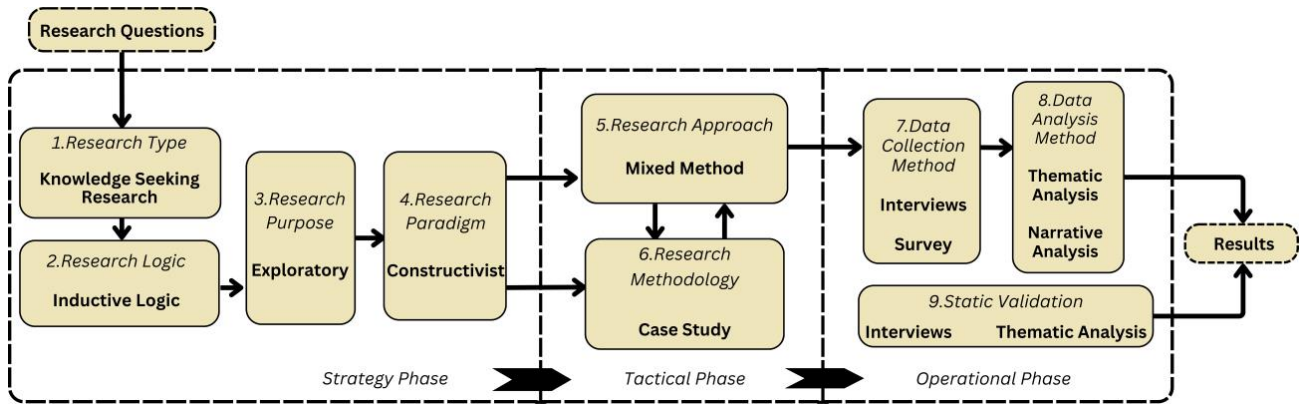

Figure 1. Decision-making structure for the research design

The research questions were developed based on the literature review and were used to guide the direction of creating the research design. The study employs knowledge-seeking research as it aims to increase the understanding of the factors and motivations that shape decision-making in responsible AI engineering. The study used inductive research logic as it tries to find concepts and patterns from the observed data to build generalizations from them. The purpose of the research is to do an exploratory study to investigate the area of ethical decision-making in responsible AI engineering, which is a field that is currently evolving and under-researched. The research design is based on the interpretivist/constructivist paradigm, utilizing both qualitative methods (interviews) and quantitative methods (surveys) to explore the research questions, as this study believes that facts are constructed by individuals, vary based on the context, and are subjective.

By combining qualitative and quantitative methods, this approach ensures a holistic and comprehensive analysis of the factors, roles, and motivations that shape decision-making in responsible AI engineering. Data collection using semi-structured interviews is an effective tool for understanding the experiences and perceptions of software professionals relating to central themes of the area of investigation, which are factors, roles, and motivations shaping decision-making [28]. By supporting the qualitative data with quantitative data from surveys, various trends can be identified across a larger population, which cannot be interpreted from interviews alone. This would further help in generalizing the findings and ensuring that the research captures a wide range of perspectives. The use of static validation on the research findings ensures the applicability of results in real real-world industry setup.

### B. Research Questions

The study employs knowledge-seeking exploratory research with RQ1 as a comparative explorative question, and RQ2 and RQ3 as descriptive explorative questions [29].

**RQ1:** *What are the main factors that distinguish the decision-making process in responsible AI development from conventional software development, and how do these factors influence the ethical and technical dimensions of AI systems?*

**RQ2:** *How do the various emerging roles within software development teams (like data engineer, data scientist, and AI engineer) impact the decision-making process in responsible AI development and contribute to the implementation of AI ethical principles?*

**RQ3:** *What are the motivations and values of individuals involved in the development of responsible AI systems, and how do these personal factors influence the decision-making processes in responsible AI development?*

RQ1 investigated the similarities and differences in the decision-making process between RAI development and CSD to identify the decision points that influence the ethical and technical aspects of AI software engineering, targeting both organizational and team-level Behavioral Software Engineering scenarios. RQ2 investigated the relevance and impact of emerging AI roles in the software development teams on the decision-making process and implementation of RAI principles. This research question targeted team-level Behavioral Software Engineering scenarios. RQ3 explored individual-level Behavioral Software Engineering scenarios, like motivations and values of individuals, to study the intricate interplay between individual factors such as motivation, personal preferences, and values in shaping the quality of the produced software through responsible engineering practices.

### C. Data Collection and Analysis

The study used surveys as a quantitative approach and interviews as a qualitative approach to collect relevant data [30]. The supplementary materials of this research study can be found at [30].

The interviews were semi-structured, where the majority of the questions were predetermined, but questions were also spontaneously asked based on the responses from the interviewee. This approach helped in maintaining a structure for the interviews while allowing freedom to explore new possible areas that arose during the conversation. The interview protocol was evaluated through a pilot interview with one industry practitioner to rectify ambiguities and to ensure clarification in the questions. There were no major changes, but some clarifications for a few questions. The sampling strategy used was convenience sampling within the industrial collaboration network of the authors. The interview participants' selection/inclusion criteria were based on their current role in the organization (roles closely associated with software development) and their experience (at least five years of overall experience).

TABLE I
INTERVIEW SUBJECT CHARACTERISTICS

| Code | Gender | Age | Years in current company | Years in current role | Job Role/Designation | Primary area of expertise | Total exp (years) | Current company size |
|---|---|---|---|---|---|---|---|---|
| P1 | Male | 37 | 11 | 1 | Head of AI Development | Software | 12 | 200 |
| P2 | Male | 32 | 2 | 2 | Data Engineering Consultant | Software & AI | 6 | 10 |
| P3 | Male | 28 | 2 | 5 | Full Stack Software Engineer | Software | 10 | 9K |
| P4 | Female | 39 | 4 | 1 | Senior Technical Specialist | Software | 11 | 2K |
| P5 | Male | 37 | 2 | 2 | Full Stack Engineer | Software | 13 | 10K |
| P6 | Female | 51 | 2 | 2 | Senior Product Manager | Software & AI | 25 | 10K |
| P7 | Male | 34 | 5 | 5 | Machine Learning Engineer | Software & AI | 8 | 10 |

Seven interviews were conducted with one interviewee and two interviewers at a time. Each interview lasted 45-75 minutes for in-depth discussion and was held in-person or via Zoom, based on participants' preference. During the interviews, the purpose of the study was presented. Then, questions about the different themes of the study were discussed in detail. All interviews were audio recorded with consent and transcribed later to facilitate and improve the analysis process. The sample size for the semi-structured interviews was decided based on the concept of data saturation, where the interviews were planned to be done until no further insights emerged [31]. Saturation was attained by the time the seventh participant was interviewed. Interview subject characteristics are shown in Table I.

Surveys were used to ascertain the findings from the interviews, and 51 responses were collected. The questionnaire consisted of a mix of closed-ended questions, including Likert scale (5-point scale) questions and open-ended (free text) questions based on the key themes of the study. The survey participants were selected through convenience sampling and snowball sampling [32]. The data collection lasted for a span of 1.5 months. The authors directly contacted connections inside the tech industry, and the questionnaires were also distributed in reputable online forums (Reddit and StackOverflow), a networking platform (LinkedIn), and professional groups focused on AI, software, and technology. Of the 51 participants, 55% were male and 40% were female. The largest segment of respondents is in the area of software development(51%), with second largest group involved in both software and AI development, making up about 27.5% of the responses. The dominant roles were Software Developer/Engineer (39.2%), Data scientist (13.7%), Quality Assurance Engineer (9.8%), and AI Software Developer/Engineer (5.9%). For full demographic information, refer [30].

Thematic analysis was used to analyze the qualitative data from interviews using Braun and Clarke's approach [33] as this method was well-suited for synthesizing insights from interviews. It involves 6 main phases namely familiarizing with data, generating initial codes, searching for themes, reviewing themes, defining and naming themes, and producing the report [33]. The analysis started with transcription of audio data, then organizing the data into sections based on interview questions, identifying patterns in the data to generate initial code, attaching these codes to sub-themes/themes, reviewing the themes to avoid duplication and finally matching the codes and themes to the relevant research questions. To ensure reliability, the codes and themes identified by two researchers working on the project were compared, and disagreements and conflicts were discussed and resolved.

To further complement thematic analysis for enhanced contextualization of the themes, to uncover themes within social realities by exploring the story telling elements and to enhance the credibility of finding to support triangulation of data, narrative analysis was used on the storytelling parts or narratives in the interviews [34], [35]. Narrative analysis examined the stories or narratives shared by individuals to understand how they construct meaning and make sense of their experiences. By understanding these narratives, researchers can uncover underlying beliefs, values, and motivations that shape individuals' perspectives and behaviors [36]. In the context of this research, narrative analysis supported the thematic analysis by revealing the underlying values, beliefs, and motivations of individuals involved in the development process of RAI systems and gained insights into a software professional's experiences of this major transformation process from CSD to RAI development. The narrative analysis involved content analysis, structure analysis, and context analysis to generate inferences from the narratives based on Murray's narrative analysis approach [37]. Contextual analysis involved finding the broader context within which the narrative is situated by understanding the social and situational factors that are influencing the stories. Content analysis involved examining the content of the stories to identify themes, patterns, and meanings to understand the underlying meanings and beliefs of the participants. Structure analysis involves finding the organization of the narrative, like the flow of the content, its coherence, and the way of organizing the stories. For enhanced clarity and comparison of qualitative data from surveys and for easy identification of trends and patterns in the responses, diverging stacked bar chart were also used for the visualizing data with both positive and negative attributes [38] and for comparing the attitudes of respondents around a specific topic [39].

To validate the finding from the above analysis, static validation of the research findings were done with four industry experts from the field of AI development. The sampling strategy used was convenience sampling within the industrial collaboration network of the authors by identifying experts in the AI field. Interviews and discussions were conducted based on the research findings to confirm its validity and applicability in an industrial setup. The selection/inclusion criteria of interview participants were based on their current job role (relevant to AI related decision making) and their knowledge in AI development decision making process (academic and industrial). Four interviews were conducted with one interviewee and one interviewer at a time. Each interview lasted 45-60 minutes for in-depth discussion and were held online. During the interviews, the purpose of the study and the research findings were presented, followed by the validation of findings through discussions. All interviews were audio recorded with consent and transcribed later in order to facilitate and improve the analysis process. Interview subject characteristics for static validation of research findings is given in Table II.

## IV. RESULTS AND ANALYSIS

### A. Thematic Analysis of interviews

The main themes identified from the thematic analysis of the data are as follows:

*1) Themes for RQ1 - Unique aspects of RAI:*

**Agility & flexibility:** Given the fast evolution in AI, strategic approaches to incorporate adaptability, agility, and modular methods in AI development are needed in AI development in comparison to CSD methods. As one participant stated, *"we take a more modular approach to developing AI software because the landscape is changing so fast"* (Participant 1). While doing so, there is a need for critical evaluation to identify the viability, ethical implications, and functional quality of AI models. In addition, uncertainty in AI development calls for a more cautious and iterative approach to decision making throughout the development cycle.

**Ethical consideration:** Prioritizing ethical aspects with a focus on data privacy, transparency, security, and biases is more essential in AI software development, though it was also relevant in CSD. This was stated by a participant, *"the first thing that comes to my mind is the data privacy of the users. Then comes transparency on how we process this data, then security, ethics, and biases"* (Participant 3). AI software development comes with more responsibility on developers and managers to take into account the complexities of data usage, privacy, and consent in AI projects, which are more pronounced in AI development compared to non-AI software development. Ethical perspectives must be embedded from early stages of an AI project with emphasis on ethical transparency and control in AI, by acknowledging the limitations of the AI software by being transparent about uncontrollable aspects in AI software, and encouraging users to apply their judgment. A participant highlighted that *"we need to build in control as much as we can, and the part you cannot control, you must be transparent about it to the users"* (Participant 6).

**Analytical and Probabilistic Nature:** Despite the overall similarity in AI development and CSD, the unpredictability of results in AI software development highlights the probabilistic nature of AI development. A participant highlighted, *"the results are sometimes not exactly what you expect, as in conventional software development."* (Participant 3). In addition, data has become a new and important decision-making aspect in AI development. A participant highlighted this by saying, *"the key difference I would say is that decision-making around data."* (Participant 2). The data not only drives the functionality of the system but also shapes the behavior of the system and its learning process and adaptability. Early decision-making is crucial in AI projects, especially in requirement engineering and design, as AI systems rely a lot on data for training and validation, so decisions regarding the sources, quality, and quantity of this data are critical due to their direct impact on the performance and accuracy of the AI models.

*2) Themes for RQ2 - Role variation in RAI:*

**Versatility and Multifaceted Roles:** Professionals working with AI software development are not limited to a single area of expertise; instead, they must engage in various roles based on the project requirements. A participant said, *"My main role is machine learning engineer, but depending on the project, it can span from just pure ad hoc data analysis to doing data engineering to build a model"* (Participant 7). This shows the need for a broad range of skill sets and to be adaptable to sustain in the field of AI development. Specialized roles like data engineers, ML engineers, and AI ethicists bring deep and diverse expertise that improves the overall development process. This will further help in building software with higher quality, as they contribute to better-informed decision-making by providing detailed insights and expert opinions. A participant stated, *"they are absolutely relevant. The umbrella of software engineering can be attributed to many different things, but we engineers have one field of expertise"* (Participant 3). This highlights the growing complexity and breadth of the software engineering field, emphasizing the critical need for specialized roles combining technical and ethical aspects to address specific challenges and tasks. Management and leadership roles in addressing the challenges and opportunities of AI integration are also relevant for facilitating strategic decision-making in responsible AI engineering.

**Collaboration and Team Dynamics:** The diverse backgrounds and experiences of team members enhance decision-making processes and facilitate open discussion and collaborative problem-solving within the team. This is very crucial for developing fair, ethical, and inclusive AI solutions. Diverse perspectives can be particularly valuable in AI projects, as biases need to be identified and mitigated. It also highlights the significance of both technical roles and roles specific to ethics in building a RAI ecosystem, where each role contributes unique perspectives and skills to navigate ethical complexities in AI software development. A participant said, *"When you are collaborating between the different disciplines, then you* have to look at the bigger picture and have discussions with more people" (Participant 7). This shows that the collective intelligence and collaboration among different roles can help to identify and mitigate long-term impacts of AI through a critical balance between technical execution and ethical foresight in AI development. With many roles having a lack of knowledge on AI implications and RAI practices, institutionalizing ethical considerations and responsible practices must be enforced regardless of roles and responsibilities within an organization.

TABLE II
INTERVIEW SUBJECT CHARACTERISTICS FROM STATIC VALIDATION INTERVIEWS

| Code | Age | Gender | Current job role | Primary area of expertise | Total experience | Current company (Size) | Years in current role | Years in current role | Current sector of work |
|---|---|---|---|---|---|---|---|---|---|
| SV1 | 37 | Woman | Data Scientist | Software & AI | 12+ years | 19000 | 5 | 5 | Tech |
| SV2 | 36 | Woman | Data Scientist | Software & AI | 2.5 years | 20000+ | 1 | 1 | Financial |
| SV3 | 47 | Man | CTO | Software & AI | 16 years | 7 | 2.5 | 2.5 | Healthcare |
| SV4 | 28 | Woman | Specialist(SE) | Software & AI | 6.4 years | 3000 | 1 | 1 | Tech |

*3) Themes for RQ3 - Human factors in RAI:*
**Personal values influencing decision-making:** Personal values like moral integrity and the desire to contribute positively to society can shape professional practices regardless of the specific role or business context. A participant stated, "*the reason I talk about ethics and so on within AI purely comes from my personal views*" (Participant 7). Personal values and proactive ethics of an individual have a transformative potential within an organization to promote organizational practices to a more ethically aligned direction. It also highlights the influences of individual ethical standards and values on professional decisions, emphasizing the role of personal morality in guiding actions and decision-making within the tech industry. This was evident from the following comment of a participant, "*I am always the first user of the software that I built. I always put myself in the position of my client, like how I want my data to be captured and used*" (Participant 3). The development philosophy of placing oneself in the user's shoes is crucial to creating ethical, user-friendly, and privacy-conscious software solutions. Ultimately, motivations that are externally imposed (e.g., legal requirements like GDPR) and those that are internally driven (e.g., the desire to build trust with users) need to be balanced to align professional and legal goals with personal ethics.

**Organizational motivators and values:** To effectively influence decision-making at the organizational level, ethical aspects need to be operationalised by translating abstract ethical principles into concrete actions and practices within the software development process. In addition, organizations must be transparent about the limitations of AI and acknowledge AI's probability of leading to errors. A participant stated, "*You should apply ethical standards so you do no harm to people and be upfront and transparent that this is an AI product and it could make mistakes*" (Participant 6). This highlights a need for continuous monitoring to manage risks effectively. It is to be noted that in smaller companies, ethical considerations might be more integral to the company culture and decision-making processes, whereas in larger companies, such issues may become less central or more bureaucratically managed, potentially reducing the focus on ethics.

B. *Narrative Analysis*

Narrative analysis was conducted on the responses for the questions where participants relied on storytelling to explain different cases and situations to portray their experiences, perceptions, and the meanings they attach to their work [30]. The key 4 themes emerged from this analysis are as follows:

*1) Differences Between Decision-making in CSD and RAI Development:* The narrative analysis showed that decision-making in RAI development is perceived as more complex and is filled with additional ethical, technical, and operational challenges compared to CSD. AI system development involves more technical challenges like compliance and interpretability of AI models when compared to conventional systems, and AI systems have more operational differences, like the need for constant monitoring after deployment and handling the unpredictable outcomes in the system. The complexity of AI software development indicated the need for ethical consideration, transparency, and awareness of societal impacts in AI development when compared to CSD. The narratives on RAI development were often framed around ethics, transparency, and the implications of AI technologies on society, which indicated that the respondents were likely influenced by current debates, industry practices, and personal experiences within the tech field.

*2) Resources and Support Needed to Better Incorporate Ethical Considerations at Work:* The narrative analysis emphasized the need for more structured support systems to address ethical concerns in AI development. This underscores the complex nature of ethics in AI and highlights the necessity for comprehensive resources like better educational resources, clearer ethical guidelines, standardized frameworks, structured checklist of ethical aspects and guidelines within the company, certifications to guide ethical practices and more practical training, research, and educational materials focused on ethics in software development. Technical resources like better datasets, error-free historical data, ethical test data, and improved data quality that reflects real-life scenarios are also essential for effectively implementing ethical considerations at the project level. Operational support from specialized roles (data scientist, AI Engineer, AI ethicist) was viewed as essential, given the complexity in ethical decision-making in AI development projects. It was seen that the narratives were mainly shaped based on the respondents' experience in the industry environment, which is faced with ethical challenges while developing AI technologies.

*3) Challenges in Implementing Ethical Principles in AI Projects:* The narrative analysis indicated that implementing ethical principles in AI projects encompasses diverse and complex challenges that could be technical, operational, or organizational. Technical challenges that affect the implementation of ethical principles include a lack of quality data, difficulty in debugging AI systems and ensuring data protection and accuracy, and post-deployment monitoring of AI systems due to data privacy issues and regulations. Operational challenges include time constraints, increased costs, and the lack of large, trustworthy datasets. Additionally organizational challenges arise from a reluctance within companies to allocate the necessary resources to focus on the ethical implications of AI development projects as well as a lack of an ethical culture within organizations in addition to challenges in balancing technological advancement with fairness, handling the implications of imposed restrictions, and the lack of understanding and awareness about ethical implications within teams or broader organizations. The narrative analysis showed that despite the listed challenges, there is a growing awareness of the ethical implications of AI development and AI technologies.

*4) Motivations and Values of Individuals within the Development Team:* The narrative analysis emphasized the significant influence of personal values on decision-making processes in software and AI development, but the real-world application of these values is often constrained by organizational factors, economic pressures, and external regulations. Thus, the insights from this analysis indicate that there is a need to encourage an ethical culture within companies and the importance of supportive structures at the managerial level to empower ethical decision-making. The personal values and motivations of individuals directly influence how decisions are made, particularly in terms of ethics, accountability, and innovation. It is to be noted that individual values can affect not just decision-making but also the quality of the output and how the development process is done. Moreover, the importance of ethics is sometimes compromised when economic benefits and external constraints like market forces and regulations are considered. The narratives were mainly shaped by the industry experience of the individuals, their personal ethics, and their knowledge about responsible development. It also shows that there is a conflict between ideal practices and practical aspects. The narrative also shows their perceptions of the role of personal values in a complex and often constrained professional environment.

*C. Analysis of survey*

*1) RQ1 - Unique aspects of RAI:* As shown in Figure 2 for the question about the sense of responsibility for ethical implications (Q1), a substantial portion of participants expressed a high sense of responsibility, indicating a strong personal commitment to ethical considerations in AI/software development. For the question regarding the impact of conflicting personal values (Q2), responses were more evenly spread over the scale. While conflicting personal values influence decision-making, the majority of respondents feel that they do so in a moderate way, indicated by the concentration in the middle. Moving to the question about organizational preparedness for RAI (Q3), respondents were more centered around levels 3 and 4, and they had a generally positive perception about how well their organization is prepared. Regarding the understanding of ethical principles (Q4), most of the respondents claimed to have a moderate and good awareness and knowledge of ethical principles. Responses to the question about the impact of regulations (Q5) showed a significant spread, with a notable amount indicating that regulations significantly facilitate their work (level 4 and 5), while others found regulations to hinder or have no substantial impact. The majority of responses were positive for the question about the decision-making process in RAI (Q6), which shows satisfaction with the decision-making processes related to responsible AI engineering. Lastly, for the question about the effectiveness of methodologies or frameworks (Q7), most responses were at level 3, with fewer at levels 4 and 5. This shows a neutral perception of the effectiveness of methodologies and frameworks, though not uniformly seen as very effective by all respondents.

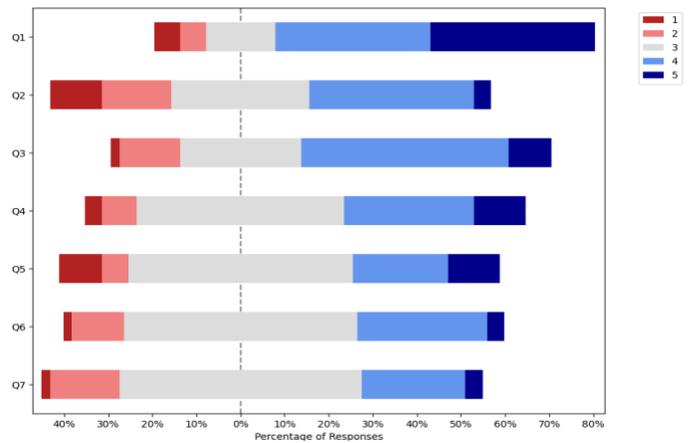

Figure 2. Diverging stacked bar chart - Likert questions for RQ1

| | |
|---|---|
| Q1 | How responsible do you feel for the ethical implications of the AI/software you develop?? (1: Not at all - 5: Completely) |
| Q2 | How much do conflicting personal values within the team affect decision-making for RAI? (1: Negligible impact - 5: Substantial impact) |
| Q3 | How prepared is your organization to integrate RAI principles? (1: Not Prepared at all - 5: Very Well Prepared) |
| Q4 | How would you rate your understanding of ethical software/AI principles? (1: Very poor - 5: Excellent) |
| Q5 | Are you aware of any AI regulations in your region, and how do they affect your work? (1: Significantly hinders - 5: Significantly facilitates) |
| Q6 | How do you rate the decision-making process in the development of RAI in your organization? (1: Very poor - 5: Excellent) |
| Q7 | How effective are the methodologies or frameworks your organization uses to ensure RAI development? (1: Not effective - 5: Very effective) |

*2) RQ2 - Role variation in RAI:* Figure 3 shows a strong endorsement of interdisciplinary collaboration in AI development. None of the respondents believes that interdisciplinary collaboration is not critical for RAI development. A very small percentage, just over 2%, of respondents view it as slightly critical. Approximately 30% view it as somewhat important, placing it at Level 3. Over two-thirds of the participants, 66.6% rated it as either important or extremely critical. For

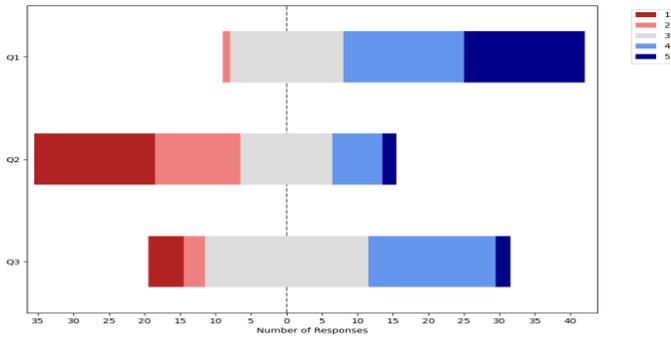

Figure 3. Diverging stacked bar chart - Likert questions for RQ2

Q1      How critical is interdisciplinary collaboration in RAI development? (1: Not critical - 5: Extremely critical)
Q2      How often do you collaborate with non-tech experts (e.g., ethics, law) in AI projects? (1: Never - 5: Always)
Q3      How effective is your organization's ethical software/AI development training? (1: Not effective - 5: Very effective)

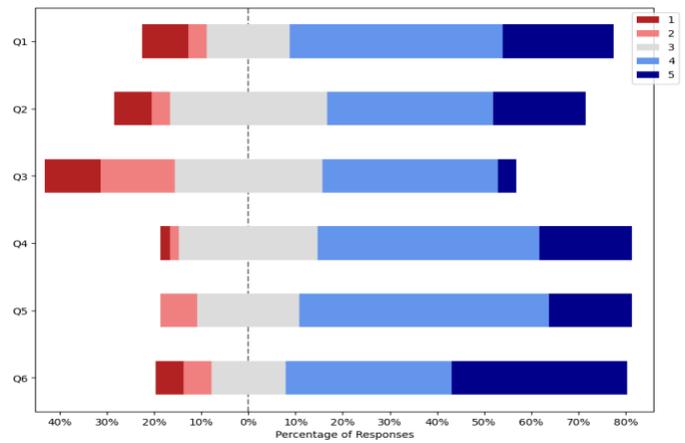

Figure 4. Diverging stacked bar chart - Likert questions for RQ3

Q1      How have your personal values influenced your approach to Software development or AI software development? (1: Not at all - 5: Extremely)
Q2      To what extent did your values significantly influence the development of a responsible software/AI system? (1: Not at all - 5: Extremely)
Q3      How much do conflicting personal values impact team decision-making in RAI development? (1: Negligible impact - 5: Substantial impact)
Q4      To what extent do your personal values align with the organizational values? (1: Not at all - 5: Completely)
Q5      How well do your ethical values in AI/software development align with your organization's ethical guidelines? (1: Not at all - 5: Completely)
Q6      To what extent do you feel responsible for the ethical implications of the AI/Software systems you develop? (1: Not at all - 5: Completely) the question about the frequency of collaboration with non-technology experts (Q2), a significant portion, around 30%, of respondents report that they never collaborate, and only nearly 4% of the respondents always collaborate with experts from non-technology fields such as ethics, sociology, or law in their AI projects. For the question about the effectiveness of training on ethical software/AI development (Q3), the responses were mostly neutral or at level 4, and more respondents rated the training as not effective (level 1) than as extremely effective (level 5).

*3) RQ3 - Human factors in RAI:* According to Figure 4, the majority of respondents for the question about personal values influencing development practices (Q1) indicated a high level of influence and considered their personal values crucial in guiding their development practices. For the question about the extent of values influencing responsible development (Q2), responses were mostly skewed toward the higher end, level 4 and 5, and this shows their personal values play an important role in responsible development practices. The impact of conflicting personal values (Q3) was claimed to be moderate. For the question about the alignment between personal and organizational values (Q4), most responses fell into the higher categories, which indicated a strong alignment between personal and organizational values. Similarly, in Q4, there is a lean towards levels 4 and 5 for the questions about the alignment of ethical values with organizational guidelines (Q5) and feeling responsible for the ethical implications (Q6).

### D. Static validation of Findings

To verify the practical relevance and applicability of the above findings in an industry scenario, interviews were conducted with four industry experts from the AI development field. The interviews confirmed the findings from RQ1 about the significant difference in AI and CSD, with AI development requiring deeper consideration on aspects like transparency, accountability, and privacy due to its high level of ethical complexity and probabilistic nature. The interviewees also favored the need for a structured ethical framework within the organization and added that the frameworks must be concrete and narrative and evolve with time. For the recommendation on post-deployment monitoring of AI software behavior, interviewees stated that it is indeed beneficial to have this continuous monitoring, but it depends on the organization's capacity to provide the resources and investment for this type of monitoring (SV1) and the type of AI projects (SV3). For RQ2, the interviewees agreed on the relevance of new roles for collaborative decision making and supported the suggestion for an H-shaped competency. They also added that H-shaped competency must not be enforced as a mandatory requirement but as a desirable trait because not all professionals with a technical background will be interested in getting trained (SV1) and becoming experts in aspects related to ethics. They acknowledge that dual expertise is possible yet challenging for technical professionals to handle both responsibilities well (SV2), but understanding basic ethical principles and knowing whom to consult during dilemmas could be practical (SV3). For RQ3, all interviewees agreed that their personal values enforce personal ethics and guide their decisions, specifically on ethical aspects like data privacy and client data confidentiality. They also emphasized the significance of organizational support, fostering a culture of ethics, empowering individuals to raise concerns about ethical issues, and targeted ethics training for fostering a culture of responsibility within the organization, thus favoring the finding on the importance of organizational culture for RAI practices. SV3 emphasized that having an ethical review board will be more feasible in larger organizations, but it would be easier to develop a culture of ethics within the organization to enforce accountability and ethical decision making.

## V. DISCUSSION

This section discusses possible rationales that can be derived from the results and findings of this study.

### A. Research Question 1 - Unique aspects of RAI

The decision-making process in RAI development significantly diverges from CSD due to unique challenges in ethical and technical aspects. Responsible AI development prioritizes ethical implications and societal impacts, while CSD often focuses on performance and functionality. The need for agility and adaptability in AI development requires strategic decision-making during the planning, design, and development stages to adapt to changing requirements. This is in line with Mendes et al. [3], who emphasized the shift towards value-based decision-making that considers long-term value creation. The development of AI software demands a deeper analytical approach to data handling, including sourcing, storage, and usage, while emphasizing ethical aspects such as accountability, transparency, explainability, and data privacy. This emphasizes that data is the primary determinant of decision points in AI software development.

Prioritizing ethical considerations involves understanding the project's requirements, the data used, and the target audience of the project. Therefore, the decision-making process also involves the prioritization of the ethical needs of the project by a thorough analysis of the project at different phases. Ethical guidelines should be iteratively tailored during the design and development phases based on the project's priorities, and the testing and validation protocols must address ethical concerns to ensure fair outcomes. Continuous monitoring of the AI system's behavior is necessary to identify and resolve ethical issues post-deployment. These findings resonate with Gogoll et al. [20], who argue that traditional codes of conduct often fail to guide complex ethical situations, highlighting the need for a systematic and structured decision-making approach that incorporates ethical and social implications. Also, MacNamara et al. [21] highlight the challenge of applying abstract ethical principles to specific decision-making contexts, indicating the need for practical, scenario-based ethical guidelines.

Due to the probabilistic nature of AI outcomes and AI's complexity, a distinct strategy is required for testing and deploying the AI software product [12]. This complexity, perceived as higher compared to CSD, may be due to professionals' limited experience with AI, as it is a newer field. This shows the importance of continuous learning and adaptation in the field, as professionals become more accustomed to AI technologies over time, potentially reducing the perceived complexity. The high sense of responsibility among professionals and their moderate to good understanding of ethical principles suggest growing awareness of AI's consequences.

The mixed views on the regulations for RAI practices indicate that there is a lack of operational frameworks, better strategies, and resources specifically dedicated to implementing more robust ethical practices within the organization, indicating that there is still room for improvement. Gogoll et al. [20] highlight the issue of under-determination, where existing codes offer inadequate guidance for specific scenarios, causing ethical ambiguities. Similarly, Heldal et al. [40] emphasize the need to integrate ethics into software engineering as a standard practice to align business practices with sustainability values. Although the focus was on differences specific to AI development, it is important to recognize that these challenges, such as ethical considerations and data handling, are inherent in both AI development and CSD. AI has more clearly highlighted these issues, which might have been less visible or often unhandled in conventional software systems. As a result, AI development has brought existing issues to the forefront, and there is a need for more structured and explicit ethical guidelines and decision-making frameworks.

This research recommends the need for ethical decision-making guidelines to effectively operationalise RSE practices for AI, where an ethical review board monitors the project life cycle of AI software development. The key recommendations include conducting thorough reviews of data sources and algorithms to identify potential biases, ensuring compliance with legal and regulatory standards, promoting transparency about data usage and AI system capabilities, and integrating ethical considerations, such as accountability, fairness, and privacy, into the design and development phases. To ensure ongoing compliance, regular stakeholder engagement and iterative ethical evaluations are also suggested, with a focus on clear documentation and feedback mechanisms for users to raise concerns and provide insights for improvements.

### B. Research Question 2 - Role variation in RAI

As stated by Zhu et al. [15], for operationalizing AI principles, it is essential to set up team members with diverse specializations and backgrounds (background, culture, discipline). This is emphasized by the findings of this study, which show that specialized roles in the field of AI development are designed to address both the technical and ethical complexities that arise from AI development, emphasizing the relevance of giving equal importance to both aspects. In contrast, the role of software engineers placed less focus on ethical considerations during the software development process. According to Schiff et al. [6], the disciplinary divide (perspectives, attitudes, values, and communication norms) between professional disciplines can hinder responsible AI engineering and governance. However, the study shows that the diverse perspectives and skill sets that are achieved from these new roles can enhance collaborative decision-making to produce more ethical and robust AI systems. Additionally, interdisciplinary collaboration of experts is encouraged to validate decision-making from multiple perspectives, particularly in the early stages of the project. To avoid possible challenges in this disciplinary divide, H-shaped competency (also known as $\pi$ shaped competency) [41] can be encouraged in professionals working with AI software development. This means that professionals should hold a high degree of competence in two distinct domains, where one domain will be related to the ethical aspects of software/AI development. This ensures that dual expertise in technical and ethical domains can help to integrate the knowledge and expertise from these two interdisciplinary fields to create responsible software.

Every stage of the RSE process for AI can integrate operational RAI principles by incorporating multiple perspectives from these specialized fields [42]. However, professionals often lack clarity about the decision-making power of various roles in AI development practices, leading to a lack of awareness about which roles have a greater impact on the decision-making process and are more relevant in the context of responsible AI engineering. This demands an organizational-level effort to make teams and individuals aware of their responsibilities in responsible AI engineering, as well as the roles and responsibilities of other team members in responsible AI engineering, promoting more effective collaboration in responsible AI engineering.

*C. Research Question 3 - Human factors in RAI*

It can be interpreted from the findings that the values and ethical considerations of individuals can critically influence the choices they make in developing software systems, as stated by Heldal et al. [40]. Integration of ethics into development practices at the organizational level can ensure more sustainable AI software. Therefore, it is justifiable to support organizational cultures that actively endorse and incorporate these shared personal and organizational values, which could result in improved productivity and more successful ethical practices in projects involving AI. As highlighted by Mendes et al. and Winter et al. [3], [24], individual values can shape the decision-making process, which was evident from the study. But personal ethical values can be disregarded due to organizational influences. This shows that external factors can affect personal motivations in a work environment. Also, it indicates that larger organizations may need to implement more robust frameworks to ensure that ethical considerations are not overshadowed by other business objectives. This finding is supported by Schiff et al. [6], who discuss the misalignment between business incentives and responsible AI engineering principles. Among the key values and principles in software/AI development, accountability and transparency were given more relevance, indicating the importance of building trust among the users by disclosing how the AI software works and how the data is used (transparency) and building safe and reliable software that is rigorously tested as the developers are accountable for the software they build (accountability).

## VI. THREATS TO VALIDITY

Threats to validity were considered in each phase of this study to understand the research execution, their consequences, the context of findings, mitigation strategies to prevent consequences, and how these factors impacted the study validity [43]. The participants may present the decision-making process of their organization in a more favorable light during interviews and surveys due to social desirability bias and the sensitive nature of the topic affecting the objectivity of the findings. To reduce this fear of repercussions, the study ensured anonymity and confidentiality of their responses such that the data would not be linked to the participants or their organizations. Data triangulation was also done by combining interviews with anonymous surveys and static validation from industry experts. However, the feedback study conducted with AI industry experts has the potential to introduce bias due to the possibility of leading questions, which could shape their responses in a particular direction. This limitation was mitigated by designing the feedback questions to be as open-ended and neutral as possible. The thematic analysis heavily relied on personal interpretation of results, which might have introduced researcher bias in the initial codes and themes, but repeated data reviews and collaborative discussions between researchers have mitigated this to a great extent. The findings may not be generalizable to all organizations or cultures due to a small sample size and limited geographic and organizational diversity in the sample. Despite the small sample size, data saturation was reached with the seventh participant of the interviews, indicating that the data collected was sufficient within the context of this study, making the study's findings robust within the sample. However, by understanding the investigated phenomena in one context, the study aids in comprehending other settings. For generalizability, the context and characteristics of the case companies in this study need to be compared with the context of interest.

## VII. CONCLUSION AND FUTURE WORK

This research explored the current state of practice in responsible AI engineering with a focus on decision-making and the integration of ethical aspects in responsible AI engineering. The study identified a gap between the state of the art and the state of practice in responsible AI engineering based on the analysis of responses from interviews and surveys. The findings highlighted the importance of the practical implementation of ethical principles in organizational practices, with the need for more strategic, interdisciplinary, and adaptable decision-making frameworks at the organizational, team, and individual levels to support RAI practices to meet the complex and evolving nature of AI software. This study also suggested a possible constraint in the wider comprehension of ethical aspects in AI development, as pressing concerns about data security and privacy may overshadow other ethical issues like transparency and accountability. This calls for action to reduce this gap through better comprehension and education on ethical aspects in responsible AI engineering, and incorporate sustainable ethical practices in the software engineering life cycle through planned, transparent, and value-based decision-making to adapt evolving AI technologies and their development. There is also a need for creating an organizational culture that prioritizes responsible and socially beneficial AI systems with an increased understanding of the professionals' role in shaping responsible software through clearer job role definitions, thereby fostering a culture of ethics and responsibility.

The findings recommend future directions for industry professionals to effectively make ethical decisions while developing AI software. The study recommends creating ethical leadership roles with H-shaped competency, encouraging continuous learning, developing operational ethical guidelines, and learning from past projects to make informed decisions. Future work should try to include a larger participant group in surveys and interviews to strengthen the findings with more focus on sustainability and social aspects of responsible AI engineering. Broadening the demographic scope, comparative research including more sectors or industries, case studies of RAI practices, document analysis of corporate decision-making strategies, and internal project reports could provide new perspectives on ethical considerations in AI development by clarifying the internal narratives and justifications for ethical decisions. There can also be longitudinal research that might offer important new perspectives on how ethical considerations in AI development change throughout time in companies.